\documentclass[twocolumn,showpacs,%
  nofootinbib,aps,superscriptaddress,%
  eqsecnum,prd,notitlepage,showkeys,10pt]{revtex4-1}

\usepackage{amssymb}
\usepackage{amsmath}
\usepackage{graphicx}
\usepackage{dcolumn}
\usepackage{hyperref}
\usepackage{multirow}

\usepackage{setspace}
\newcolumntype{d}[1]{D{.}{.}{#1} }

\begin{document}

\title{Frontier Orbitals and Quasiparticle Energy Levels in Ionic Liquids}
\author{J. Matthias Kahk}
\affiliation{Department of Materials, Imperial College London, South Kensington, London SW7 2AZ, United Kingdom}
\author{Ivar Kuusik}
\affiliation{Institute of Physics, University of Tartu, W. Ostwaldi 1, 50411 Tartu, Estonia}
\author{Vambola Kisand}
\affiliation{Institute of Physics, University of Tartu, W. Ostwaldi 1, 50411 Tartu, Estonia}
\author{Kevin R. J. Lovelock}
\affiliation{School of Chemistry, Food and Pharmacy, University of Reading, Reading, RG6 6AT UK}
\author{Johannes Lischner}
\email{j.lischner@imperial.ac.uk}
\affiliation{Department of Physics and Department of Materials, and the Thomas Young Centre for Theory and Simulation of Materials, Imperial College London, London SW7 2AZ, United Kingdom}

\begin{abstract}
Room temperature ionic liquids play an important role in many technological applications and a detailed understanding of their frontier molecular orbitals is required to optimize interfacial barriers, reactivity and stability with respect to electron injection and removal. In this work, we calculate quasiparticle energy levels of ionic liquids using first-principles many-body perturbation theory within the GW approximation and compare our results to various mean-field approaches, including semilocal and hybrid density-functional theory and Hartree-Fock. We find that the mean-field results depend qualitatively and quantitatively on the treatment of exchange-correlation effects, while GW calculations produce results that are in excellent agreement with experimental photoelectron spectra of gas phase ion pairs and ionic liquids. These results establish the GW approach as a valuable tool for understanding the electronic structures of ionic liquids.
\end{abstract}

\maketitle

\section{Introduction}

Room temperature ionic liquids (RTILs) are salts formed of molecular cations and anions that exist in the liquid state at or near room temperature. They find widespread use as solvents,\cite{rogers_chemistry_2003,welton_room-temperature_1999} dispersants,\cite{boukherissa_ionic_2009,wu_functionalization_2009,zhou_dispersion_2010} and electrolytes,\cite{balducci_high_2007,buzzeo_non-haloaluminate_2004,galinski_ionic_2006} and exhibit several unusual properties, including high electrochemical stability windows\cite{van_aken_formulation_2015,hayyan_investigating_2013} and very low equilibrium vapour pressures.\cite{ludwig_we_2007,m_s_s_esperanca_volatility_2010} From a fundamental point of view, it is important to understand the character of the frontier molecular orbitals and determine their quasiparticle energy levels in RTILs, as these determine technologically important properties such as band alignment at interfaces, reactivity, and stability with respect to electron injection or removal. In recent years, a number of experimental and theoretical investigations of the electronic structure of ionic liquids have been reported. For example, photoelectron spectroscopy has been used to study the valence band electronic structure of liquid RTILs and ionic liquid vapours consisting of neutral cation-anion pairs.\cite{fogarty_electron_2019,kuusik_valence_2016,kuusik_valence_2018,kuusik_valence_2019,kuusik_electronic_2019,lovelock_photoelectron_2010,ulbrich_photoelectron_2014}

Computational studies of RTILs have mostly been carried out in the framework of density functional theory (DFT).\cite{fogarty_electron_2019,ong_electrochemical_2011,reinmoller_theoretical_2011,ulbrich_photoelectron_2014} Advantages of DFT include its relatively modest computational cost and its ability to predict ground state geometries with good accuracy, as long as dispersion interactions are taken into account.\cite{grimme_performance_2012} DFT is also often used to gain insights into the electronic structures of materials by comparing Kohn-Sham (KS) eigenvalues to measured photoelectron spectra. However, KS eigenvalues cannot be rigorously interpreted as quasiparticle energies (with the exception of the energy of the highest occupied molecular orbital (HOMO)~\cite{janak_proof_1978}) which are measured in photemission spectroscopy. This is the origin of the famous band gap problem of DFT.\cite{perdew_density_1985} True quasiparticle energies can be obtained from Green's function techniques, such as the GW approach. In the GW approach, the one-electron Green's function $G$ is obtained by solving the Dyson equation with a self-energy which is given by the product of the Green's function and the screened interaction $W$. In principle, the GW self-energy should be evaluated using the fully interacting Green's function and screened interaction. In practice, however, a mean-field Green's function $G_0$ and a mean-field screened interaction $W_0$ obtained from a DFT or Hartree-Fock (HF) calculation are often used. This approximation, termed G0W0, has been demonstrated to produce highly accurate quasiparticle energies for a wide range of materials. For example, previous work has shown that G0W0 calculations can predict band gaps in solids and first ionization energies of small molecules with high accuracy.\cite{hybertsen_first-principles_1985,hybertsen_electron_1986,louie_first-principles_1998,bruneval_benchmarking_2013,caruso_benchmark_2016,sharifzadeh_quantitative_2012,onida_electronic_2002} Similarly, G0W0 yields accurate results for the position of the d-bands in noble metals relative to the Fermi level,\cite{bernardi_theory_2015,marini_quasiparticle_2001,yi_quasiparticle_2010} whereas standard DFT functionals do not. A downside of the G0W0 method is that the results can depend on the mean-field starting point. To overcome this problem, partially and fully self-consistent GW schemes have been introduced.\cite{vanschilfgaarde_quasiparticle_2006,caruso_unified_2012,vlcek_simple_2018}

In this work, the GW method is used to study the electronic structures of room temperature ionic liquids (RTILs). As a case study, the electronic structure of the 1-Ethyl-3-methylimidazolium tetrafluoroborate ([EMIM][BF$_4$]) ion pair is analyzed in detail with a focus on the nature of the frontier molecular orbitals in this system. Calculated quasiparticle energies from G0W0 calculations are also compared against recent photoemission measurements of several different ionic liquids. In particular, gas phase spectra of ionic liquid vapours are compared against simulated spectra of free ion pairs, and liquid phase spectra of RTILs are compared against theoretical calculations of periodic crystalline RTILs. In all cases, excellent agreement between measured photoemission spectra and GW calculations is found, while DFT results depend sensitively on the treatment of exchange-correlation effects. 

\section{Results}

\begin{figure}
	\centering
	\includegraphics{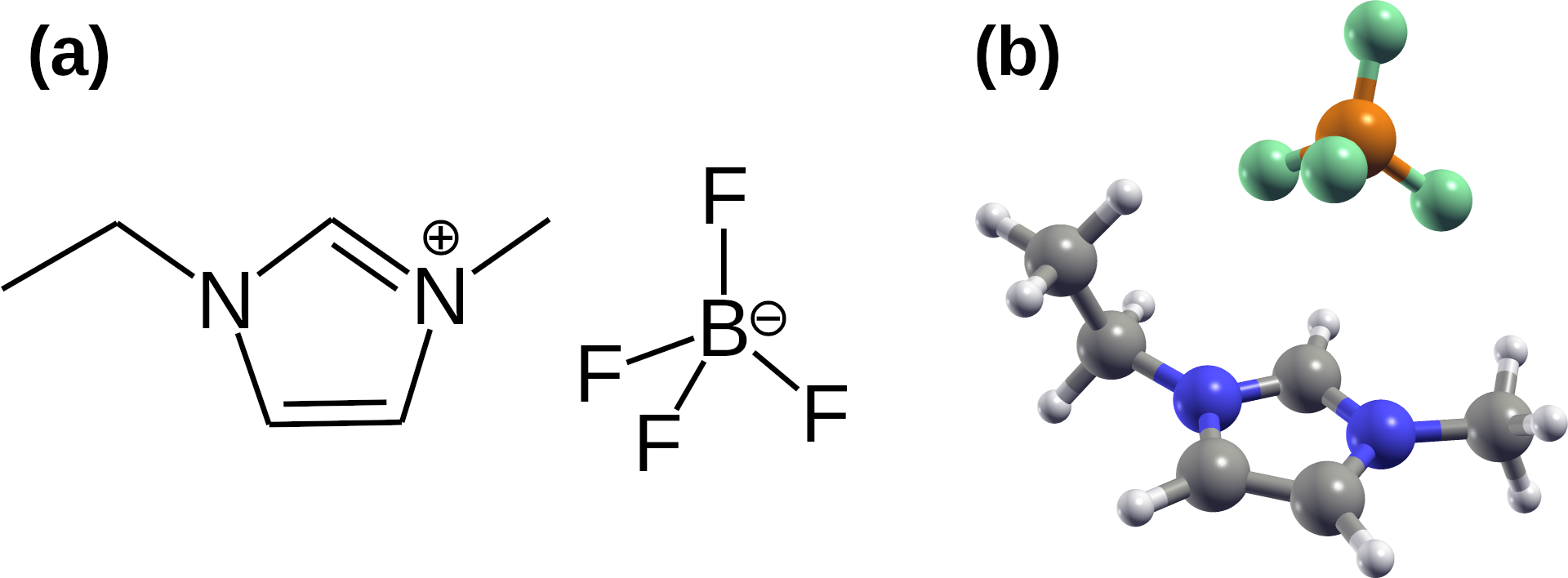}
	\caption{Structure of the [EMIM][BF$_4$] ion pair. (a) skeletal formula, (b) ball-and-stick model.}
	\label{Fig_ion_pair_structure}
\end{figure}

\begin{figure*}
	\centering
	\includegraphics{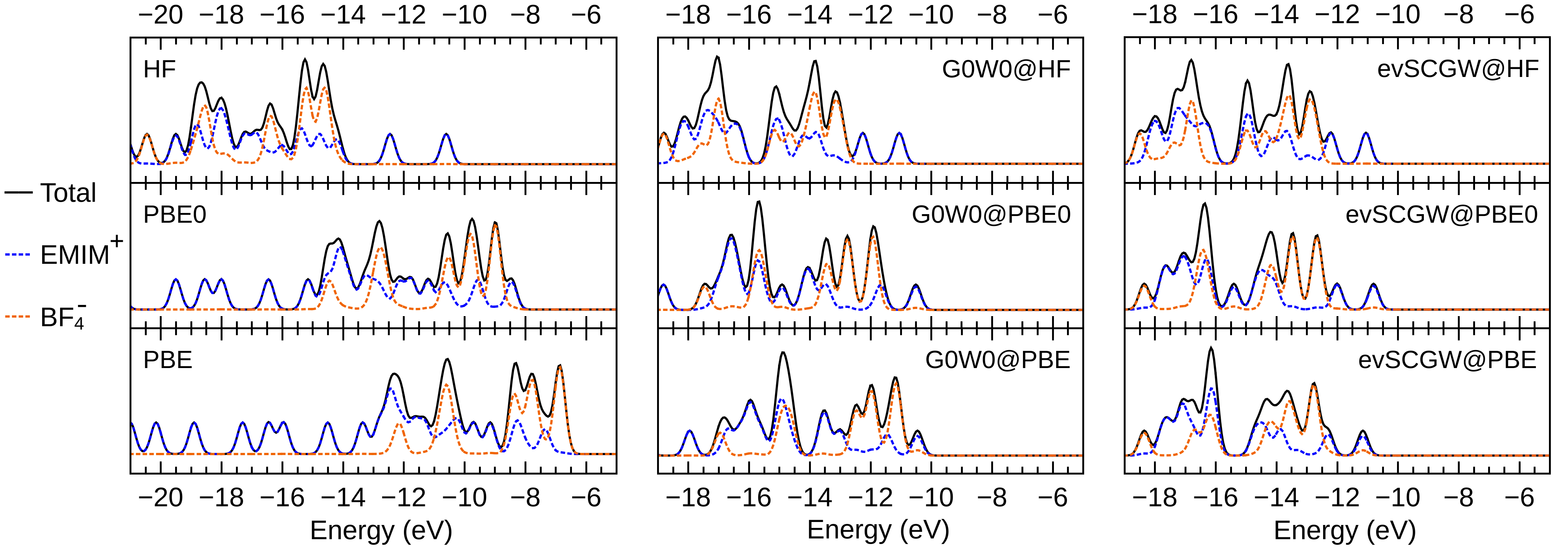}
	\caption{Densities of states of the [EMIM][BF$_4$] ion pair calculated at different levels of theory. The anion and cation contributions have been determined from a Mulliken analysis.}
	\label{Fig_ion_pair_DOS}
\end{figure*}

\begin{figure*}
	\centering
	\includegraphics[width=7.01in]{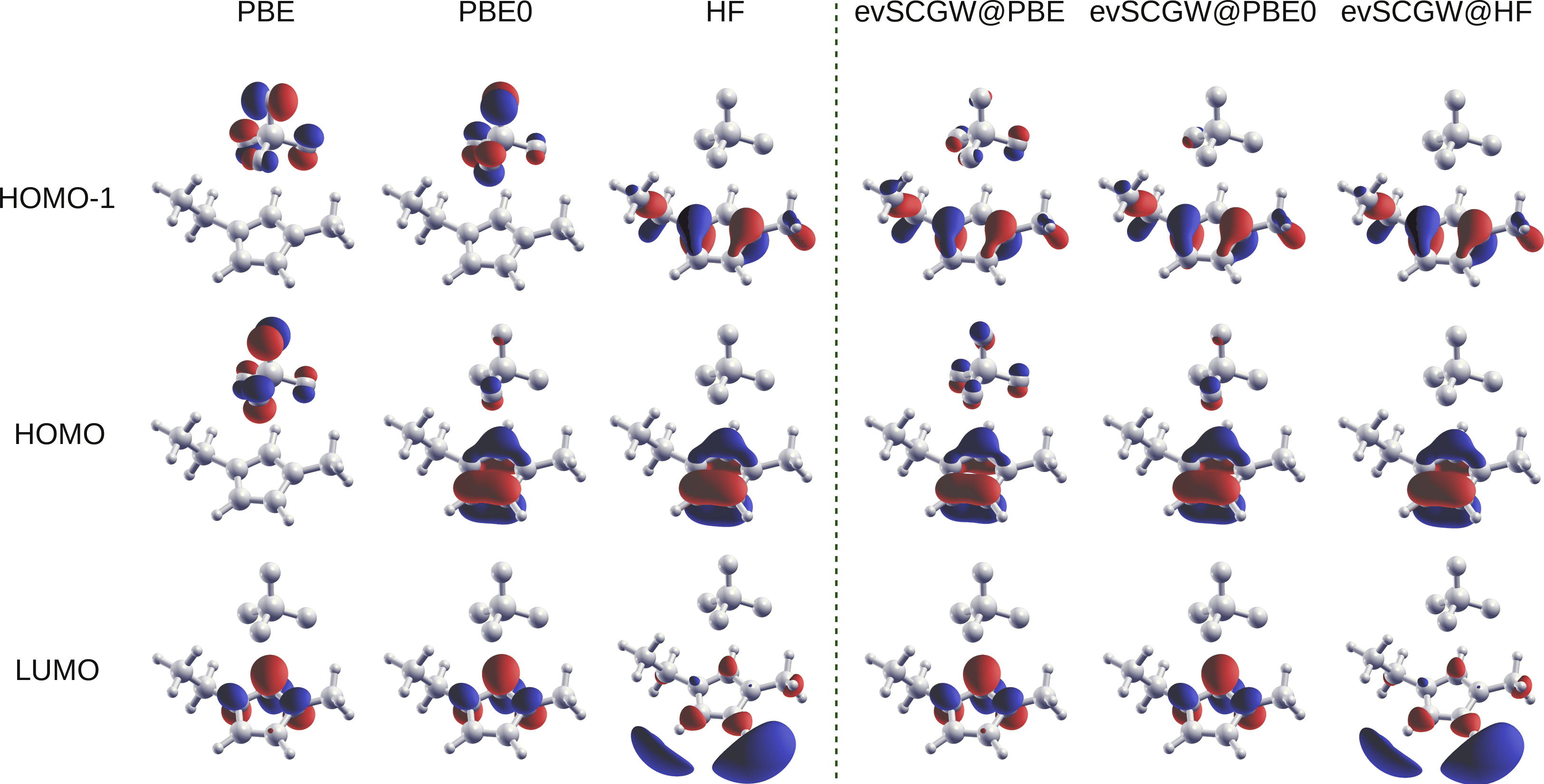}
	\caption{Frontier orbitals of the [EMIM][BF$_4$] ion pair calculated at different levels of theory. The differences between PBE (PBE0) and evSCGW@PBE (evSCGW@PBE0) arise because the energy ordering of the orbitals changes when their eigenvalues are recalculated using the GW method. The one-electron wavefunctions themselves are not updated. When Hartree-Fock theory is used as the starting point for evSCGW calculations, the energy ordering of the frontier orbitals of the [EMIM][BF$_4$] ion-pair does not change.}
	\label{Fig_frontier_orbitals}
\end{figure*}

We first consider the electronic structure of the [EMIM][BF$4$] ion pair (Fig.~\ref{Fig_ion_pair_structure}). Fig.~\ref{Fig_ion_pair_DOS} shows the calculated densities of states (DOS) of the [EMIM][BF$4$] ion pair from different levels of theory. The leftmost column contains results from three different mean-field methods: Hartree-Fock, DFT with the PBE0 functional,\cite{adamo_toward_1999} and DFT with the PBE functional.\cite{perdew_generalized_1996} The Mulliken decomposition\cite{mulliken_electronic_1955} of the total DOS into cation and anion contributions is also shown. The three curves exhibit significant quantitative and qualitative differences. For example, whilst PBE predicts that the HOMO orbital is centered on the anion, PBE0 and HF place the HOMO orbital on the cation and the associated HOMO energies differ by several electron volts among the different approaches. To illustrate this point further, isosurface plots of the HOMO-1, HOMO, and LUMO orbitals are shown in Fig.~\ref{Fig_frontier_orbitals}. The three leftmost columns show that PBE, PBE0 and HF predict three different sets of frontier orbitals in this system. In particular, all three frontier orbitals are localzied on the [EMIM] ion in HF, while the HOMO-1 in PBE0 and PBE is on the [BF$_4$] ion. In PBE, the HOMO is also localized on the [BF$_4$]. These results clearly demonstrate that standard mean-field methods are not able to unambigiously answer questions about the nature and energies of the frontier molecular orbitals in the [EMIM][BF$_4$] ion pair.

Calculated densities of states from G0W0 and eigenvalue self-consistent GW (evSCGW) calculations are shown in the middle and rightmost columns of Figure~\ref{Fig_ion_pair_DOS}. Already at the G0W0 level, the dependence on the mean-field starting point is significantly reduced and all G0W0 results 
predict that the HOMO orbital lies on the [EMIM] cation. The starting point dependence is even weaker in the evSCGW results. The frontier orbitals from the GW calculations are shown in the rightmost three columns of Fig.~\ref{Fig_frontier_orbitals} and are in qualitative agreement with each other. In particular, all frontier orbitals are localized on the [EMIM] ion. Note that the mean-field wavefunctions are not updated in either G0W0 or evSCGW, explaining the different shapes of the LUMO state of evSCGW@HF compared to evSCGW@PBE and evSCGW@PBE0. Instead, changes in the frontier orbitals arise due to changes in the energy ordering of the one-electron eigenstates when the eigenvalues are recalculated using the GW method. In [EMIM][BF$_4$], G0W0 and evSCGW change the ordering of the frontier orbitals when using a PBE or PBE0 starting point, but not when using a HF starting point. 

It is also instructive to consider the absolute energy levels of the frontier orbitals. The calculated energies of the HOMO and the LUMO of the [EMIM][BF$_4$] ion pair from different levels of theory are given in Table \ref{Table_EMIM_BF4_ion_pair}. The HOMO energies from different mean-field approaches differ by almost 4~eV with HF giving the lowest value (-10.60~eV) and PBE giving the highest (-6.80~eV). This spread is significantly reduced by the one-shot G0W0 correction with G0W0@HF still giving the lowest value (-11.03~eV) and G0W0@PBE giving the highest (-10.44~eV). Eigenvalue self-consistency does not change the G0W0@HF result, but shifts the G0W0@PBE result down by 0.6~eV. Considering next the LUMO level, we find that HF predicts a positive LUMO energy and therefore an unbound state, while the LUMO is bound in PBE and PBE0. G0W0 and evSCGW calculations confirm that the LUMO level indeed lies above the vacuum level and is unbound. Finally, we also compare the mean-field and GW results to $\Delta$-self-consistent-field ($\Delta$SCF) calculations, see Table~\ref{Table_EMIM_BF4_ion_pair}. The $\Delta$SCF method has been previously used to predict electrochemical stability windows in ionic liquids.\cite{ong_electrochemical_2011,karu_predictions_2016} We find that in the [EMIM][BF$_4$] ion pair, like G0W0 and evSCGW, $\Delta$SCF calculations predict that the LUMO lies above the vacuum level. If PBE0 is used as the mean field theory, G0W0, evSCGW and $\Delta$SCF yield similar results for both the HOMO and the LUMO energies. This agreement indicates that PBE0 is a reliable mean-field starting point for GW calculations in these systems.

\begin{table}
	\begin{tabular}{ c d{1} d{1} d{1} }
		\hline 
		\multirow{2}{*}{Method} & \multicolumn{1}{c}{HOMO} / & \multicolumn{1}{c}{LUMO} / & \multicolumn{1}{c}{\multirow{2}{*}{Gap}} \\
		& \multicolumn{1}{c}{- 1st IE} & \multicolumn{1}{c}{-E.A.} & \\ 
		\hline 
		\noalign{\vskip 1mm} 
		PBE & \multirow{1}{*}{-6.80} & \multirow{1}{*}{-1.99} & \multirow{1}{*}{4.81} \\
		PBE0 & \multirow{1}{*}{-8.44} & \multirow{1}{*}{-1.01} & \multirow{1}{*}{7.43} \\
		HF & \multirow{1}{*}{-10.60} & \multirow{1}{*}{1.89} & \multirow{1}{*}{12.49}\\
		\noalign{\vskip 1mm} 
		G0W0@PBE & \multirow{1}{*}{-10.44} & \multirow{1}{*}{0.23}& \multirow{1}{*}{10.67}\\
		G0W0@PBE0 & \multirow{1}{*}{-10.51} & \multirow{1}{*}{0.49} & \multirow{1}{*}{11.00}\\
		G0W0@HF & \multirow{1}{*}{-11.03} & \multirow{1}{*}{1.28} & \multirow{1}{*}{12.31} \\
		\noalign{\vskip 1mm} 
		evSCGW@PBE & \multirow{1}{*}{-11.14} & \multirow{1}{*}{0.80} & \multirow{1}{*}{11.94}\\
		evSCGW@PBE0 & \multirow{1}{*}{-10.79} & \multirow{1}{*}{0.79} & \multirow{1}{*}{11.58} \\
		evSCGW@HF & \multirow{1}{*}{-11.03} & \multirow{1}{*}{1.23} & \multirow{1}{*}{12.26} \\
		\noalign{\vskip 1mm} 
		$\Delta$SCF@PBE & \multirow{1}{*}{-9.55} & \multirow{1}{*}{0.43} & \multirow{1}{*}{9.98} \\
		$\Delta$SCF@PBE0 & \multirow{1}{*}{-10.20} & \multirow{1}{*}{0.58} & \multirow{1}{*}{10.78} \\
		$\Delta$SCF@HF & \multirow{1}{*}{-9.10} & \multirow{1}{*}{1.43} & \multirow{1}{*}{10.53} \\
		\noalign{\vskip 1mm} 
		\hline
	\end{tabular}
	\caption{HOMO and LUMO levels of the [EMIM][BF$_4$] ion pair from different levels of theory. All energies are given in eV.}
	\label{Table_EMIM_BF4_ion_pair}
\end{table}

We next compare GW results for different ion pairs to experimental photoelectron spectra of ionic liquid vapours. The simulated spectra are constructed from G0W0 calculations with a PBE0 starting point based on the ``Gelius approximation", i.e. the spectrum is a sum of atomic orbital projected density of states (pDOS) curves, each weighted by the per-electron photoionization cross-section of that subshell at the relevant photon energy.\cite{gelius_esca_1972} Uniform Gaussian broadening has been applied to each theoretical spectrum. 

\begin{figure*}
	\centering
	\includegraphics{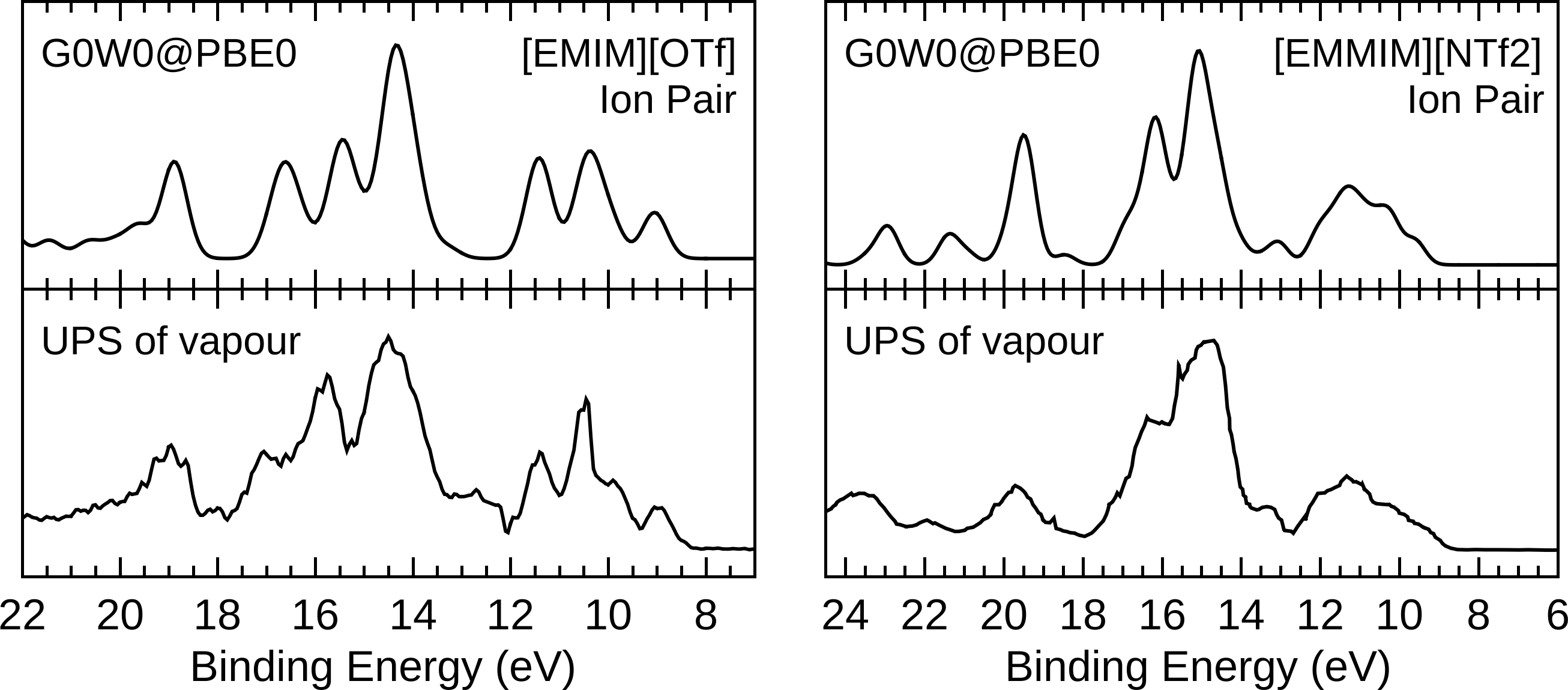}
	\caption{Valence level photoelectron spectra of free ion pairs. Theoretical results are compared against experimental gas phase ultraviolet photoelectron spectra (UPS) of ionic liquid vapours from [\citenum{kuusik_notitle_2020}].}
	\label{Fig_gas_phase}
\end{figure*}

Experimental and theoretical gas phase spectra of the 1-Ethyl-3-methylimidazolium trifluoromethanesulfonate ([EMIM][OTf]) and 1-Ethyl-2,3-dimethylimidazolium bis(trifluoro-methylsulfonyl)imide ([EMMIM][NTf$_2$]) ion pairs are shown in Figure~\ref{Fig_gas_phase}. In both cases, excellent agreement between theory and experiment is observed. We emphasize that no shifts or calibrations of any kind have been applied to the theoretical spectra, i.e. both the absolute and the relative binding energies of valence electrons in these ion pairs are predicted with excellent accuracy by the G0W0@PBE0 approach. 

\begin{figure}
	\centering
	\includegraphics{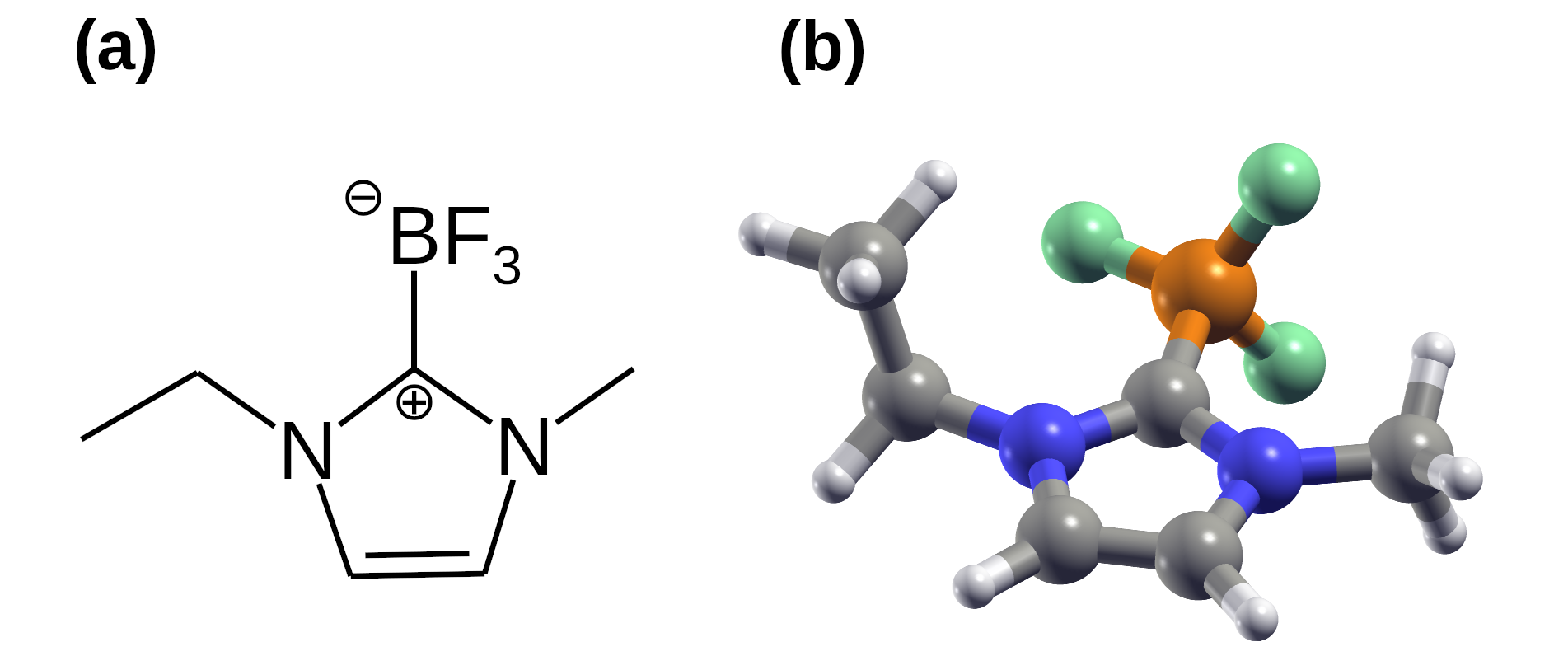}
	\caption{Structure of the ``adduct'' formed upon heating liquid [EMIM][BF$4$]. (a) skeletal formula, (b) ball-and-stick model.}
	\label{Fig_adduct}
\end{figure}

\begin{figure}
	\centering
	\includegraphics{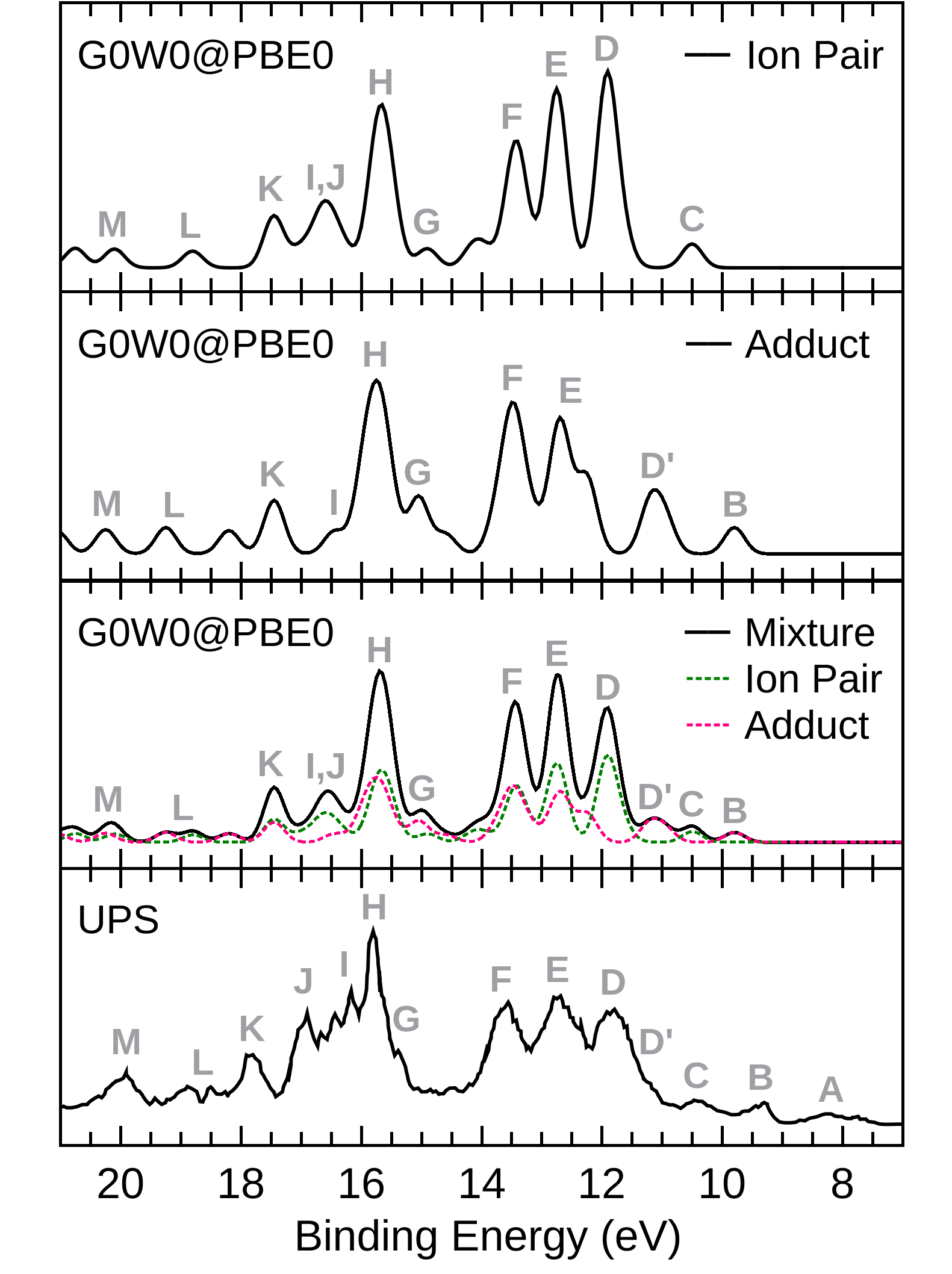}
	\caption{The experimental valence band photoelectron spectrum of the vapor above [EMIM][BF$_4$] compared against theoretical spectra. The calculated spectra of the ion pair, the adduct, and a 1.5:1 ion-pair:adduct mixture are shown.}
	\label{Fig_gas_phase_expt_vs_theory}
\end{figure}

Interestingly, the agreement between the experimental photoelectron spectrum of vaporized [EMIM][BF$_4$] and the G0W0@PBE0 result for the free ion pair is somewhat worse than for the ion pairs discussed above, see Figure~\ref{Fig_gas_phase_expt_vs_theory}. In particular, peaks A, B, and D' are missing from the simulated spectrum, and the intensity ratios of peaks D, E, and F are different from the experimental ones. In previous studies, it has been observed that the two ions of the [EMIM][BF$_4$] ion pair can react to form an adduct upon heating.\cite{clarke_thermal_2018,taylor_borane-substituted_2011} Figure~\ref{Fig_adduct} shows the structure of the adduct. To assess if adduct formation is responsible for the differences between the simulated and the measured spectra, we performed GW calculations on the adduct. We then added the adduct spectrum to the ion pair spectrum assuming that the vapour is a 1.5:1 mixture of ion pairs and adducts. Figure~\ref{Fig_gas_phase_expt_vs_theory} shows that the resulting spectrum is in much better agreement with the measurement. In particular, peaks B and D' are present and the intensity ratios of peaks D, E, and F are correct, but peak A is still missing. Including eigenvalue self-consistency or using a different mean-field starting point was also not found to reproduce peak A; see the Supplementary Information. We therefore hypothesize that this missing peak originates from a different decomposition product or an ion pair dimer.

\begin{figure*}
	\centering
	\includegraphics{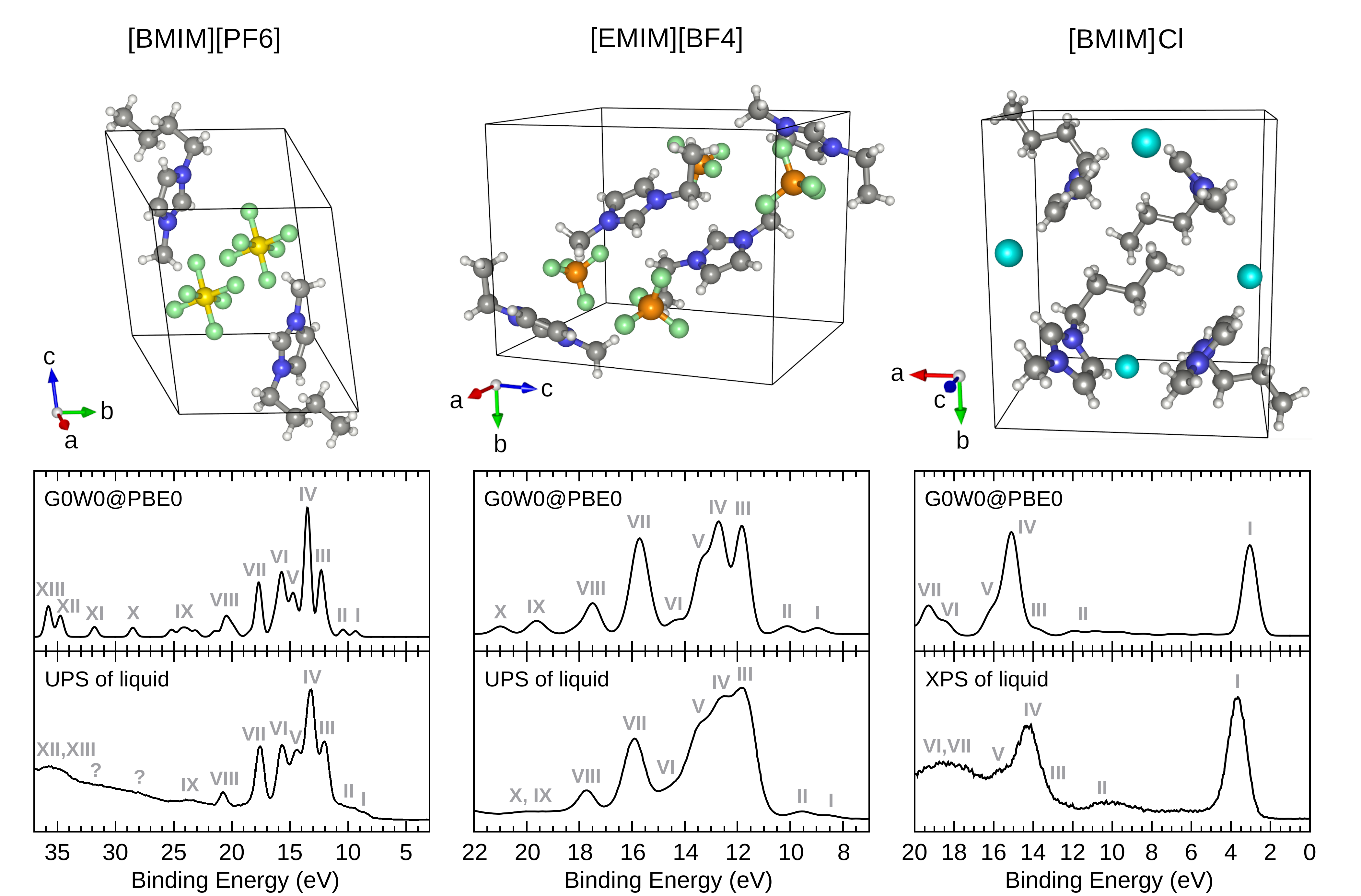}
	\caption{Calculated valence level photoelectron spectra of crystalline ionic liquids compared against experimental liquid phase photoemission measurements. The experimental spectra of [BMIM][PF$_6$], [EMIM][BF$_4$] and [BMIM]Cl have been taken from [\citenum{yoshimura_electronic_2005}], [\citenum{kuusik_valence_2018}] and [\citenum{ulbrich_photoelectron_2014}] respectively. (Digitized data is plotted.)} 
	\label{Fig_solid_and_liquid}
\end{figure*}

Finally, we also carry out GW calculations of ionic liquids in the condensed phase and compare them to experimental photoelectron spectra. In principle, the simulated spectrum of the ionic liquid should be obtained by averaging results of different liquid configurations. However, performing many GW calculations of large unit cells is computationally extremely challenging. Instead, we instead carry out GW calculations on ionic liquids in a solid, crystalline phase. This approximation is justified as the internal structures of the ions and their average coordination environments are similar in the solid and the liquid phases. Figure~\ref{Fig_solid_and_liquid} shows the unit cells of the three crystalline ionic liquids 1-Butyl-3-methylimidazolium hexafluorophosphate ([BMIM][PF$_6$]), [EMIM][BF$_4$], and 1-Butyl-3-methylimidazolium chloride ([BMIM]Cl) and also compares the simulated G0W0@PBE0 spectra to experimental photoelectron spectra taken in the liquid phase. In photoemission measurements of liquids and solids, the experimental binding energies are given relative to the Fermi level, but since the position of the Fermi level relative to the band edges is not known \textit{a priori}, the calculated spectra have been shifted by a constant amount to best match the experiment. Excellent agreement between theory and experiment is found for [BMIM][PF$_6$] and [EMIM][BF$_4$]. In the case of [EMIM][BF$_4$], peaks I and II are correctly reproduced, which is an improvement over previous DFT results\cite{kuusik_valence_2018}. In the spectrum of [BMIM]Cl, the separation between the two most intense peaks (peaks I and IV) is overestimated by approximately 1.3 eV, but otherwise the measured spectrum is reproduced with good accuracy. The results shown in Figures~\ref{Fig_gas_phase} and \ref{Fig_solid_and_liquid} demonstrate that the GW method is very well suited for predicting quasiparticle energy levels in ionic liquids and free ion pairs.

\section{Discussion}

An alternative method for modelling photoelectron spectra of ionic liquids based on DFT was proposed in reference [\citenum{yoshimura_electronic_2005}]. In this study, it was shown that experimental spectra of liquid RTILs can be reconstructed from DFT partial density of states (pDOS) curves of free ion pairs by shifting the cation and anion pDOS curves relative to each other by an amount that is determined on a case-by-case basis. The size of these shifts was originally interpreted as the difference between the average electrostatic potentials experienced by the cation and the anion. Our GW results, however, suggest that this interpretation needs to be revised. In particular, Figure~\ref{Fig_ion_pair_DOS} shows that the GW self-energy corrections give rise to a significant relative shift of the anion and cation pDOS curves. This shift does not arise from electrostatic effects, but instead from a more accurate treatment of exchange and correlation effects. Therefore, the shifts applied in reference [\citenum{yoshimura_electronic_2005}] do not arise solely from electrostatic effects and should be interpreted as empirical corrections that contain contributions from both self-energy effects and changes in average electrostatic potential.

Detailed knowledge of the character of the frontier molecular orbitals and their quasiparticle energies is of crucial importance for understanding the electronic structures of ionic liquids. In this study we have shown that interpreting DFT Kohn-Sham eigenvalues as true quasiparticle energies can lead to qualitatively and quantitatively inaccurate results. This problem can be overcome by the GW method which produces results that are in excellent agreement with state-of-the-art photoemission data. These results suggest that the GW method is a useful tool for studying the electronic structure of ionic liquids and can be used to gain insights into electronic properties that are relevant to ionic liquid devices, such as band alignment at interfaces and stability with respect to electron injection and removal.

 \section{Methods}

All calculations reported in this work were performed using the FHI-aims electronic structure program,\cite{blum_ab_2009,havu_efficient_2009,levchenko_hybrid_2015} that uses atom-centered local basis functions defined on a numerical grid. The geometries of the free ion pairs were relaxed using DFT with the PBE0 exchange-correlation functional until the forces on the atoms were less than 0.005 eV/\AA. Van der Waals interactions were accounted for using the Tkatchenko-Scheffler method.\cite{tkatchenko_accurate_2009} For each ion pair, a number of different configurations were manually constructed, and in the end the relaxed geometry with the lowest energy was used for the density of states calculations. The default "tight" numerical basis sets were used during the geometry optimizations. For the bulk crystals, the calculations were performed at experimental geometries from X-ray crystallography.\cite{choudhury_situ_2005,holbrey_crystal_2003} The implementation of the GW method in FHI-aims is described in reference [\citenum{ren_resolution--identity_2012}]. The self-energy was calculated on the imaginary frequency axis with 100 frequency points, and the Pade approximation with 16 fitting parameters was used for the analytical continuation of the self-energy onto the real axis. For the G0W0 and evSCGW calculations, the NAO-VCC-nZ basis sets were used (NAO-VCC-4Z for the ion pairs and NAO-VCC-3Z for the bulk solids).\cite{zhang_numeric_2013} All of the occupied and empty electronic states spanned by the basis sets were included in the GW calculations. A graph showing basis set convergence is included in the Supplementary Information. The bulk calculations were performed at the Gamma point only. 

\section{Data availability}

The structures of all of the ion pairs and solids considered in this work are given in the supplementary information.

\section{Acknowledgements}

J.M.K. and J.L. acknowledge support from EPRSC under Grant No. EP/R002010/1 and from a Royal Society University Research Fellowship (URF$\backslash$R$\backslash$191004). Via J.L.'s membership of the UK's HEC Materials Chemistry Consortium, which is funded by EPSRC (EP/L000202), this work used the ARCHER UK National Supercomputing Service. 
I.K. and V.K. acknowledge Estonian Centre of Excellence
in Research project "Advanced materials and high-technology devices
for sustainable energetics, sensorics and nanoelectronics" TK141
(2014-2020.4.01.15-0011).

\section{Competing interests}

The authors declare no competing interests.

\bibliography{Lllibrary_v5}

\bibliographystyle{ieeetr}

\end{document}